\pdfoutput=1
\documentclass{acm_proc_article-sp}

\usepackage{amsmath}
\usepackage{amssymb}
\usepackage{graphicx}
\usepackage{dcolumn}
\usepackage{bm}
\usepackage{url}

\newcommand{\rar}{\rightarrow}
\newcommand{\ra}{\rangle}
\newcommand{\la}{\langle}
\newcommand{\ttt}{\texttt}
\newcommand{\mbf}{\mathbf}
\newcommand{\mbb}{\mathbb}
\newcommand{\mca}{\mathcal}

\begin{document}

\title{A Recommender System to Support the \\ Scholarly Communication Process}

\numberofauthors{4} 
\author{
\alignauthor
Marko A. Rodriguez\\
       \affaddr{Knowledge Reef Systems Inc.}\\
       \affaddr{Santa Fe, New Mexico 87501}\\
       \email{marko@k-reef.com}
\alignauthor
David W. Allen\\
       \affaddr{Knowledge Reef Systems Inc.}\\
       \affaddr{Santa Fe, New Mexico 87501}\\
       \email{dave@k-reef.com}
\and
\alignauthor
Joshua Shinavier\\
       \affaddr{Knowledge Reef Systems Inc.}\\
       \affaddr{Santa Fe, New Mexico 87501}\\
       \email{josh@k-reef.com}
\alignauthor
Gary Ebersole\\
       \affaddr{Knowledge Reef Systems Inc.}\\
       \affaddr{Santa Fe, New Mexico 87501}\\
       \email{gary@k-reef.com}}
       
\date{20 February 2009}

\maketitle
\begin{abstract}
The number of researchers, articles, journals, conferences, funding opportunities, and other such scholarly resources continues to grow every year and at an increasing rate. Many services have emerged to support scholars in navigating particular aspects of this resource-rich environment. Some commercial publishers provide recommender and alert services for the articles and journals in their digital libraries. Similarly, numerous noncommercial social bookmarking services have emerged for citation sharing. While these services do provide some support, they lack an understanding of the various problem-solving scenarios that researchers face daily. Example scenarios, to name a few, include when a scholar is in search of an article related to another article of interest, when a scholar is in search of a potential collaborator for a funding opportunity, when a scholar is in search of an optimal venue to which to submit their article, and when a scholar, in the role of an editor, is in search of referees to review an article. All of these example scenarios can be represented as a problem in information filtering by means of context-sensitive recommendation. This article presents an overview of a context-sensitive recommender system to support the scholarly communication process that is based on the standards and technology set forth by the Semantic Web initiative.
\end{abstract}



\section{Introduction}

The purpose of a recommender system is to help fulfill the resource requirements of the individual enacting the service. For resource-rich environments, recommender systems serve as an information filtering tool that reduces the search space to some human-mangeable subset \cite{recommned:resnick1997}. Within this subset, the individual is better able to identify those resources that meet their current requirements. However, understanding the current requirements of an individual is a difficult problem. For example, there are issues surrounding recommendations based soley on usage data. While a particular resource was necessary in the past (e.g.~the ``Elmo's ABC book" was needed for a colleague's child's birthday), it may not be a good predicator of future requirements (e.g.~the child's birthday has passed). Thus, the problem of recommendation is exacerbated by the fact that the requirements of an individual fluctuates according to their \textit{context} (i.e~their ever-changing resource requirements).

Members of the scholarly community face a variety of contexts each with different resource requirements \cite{multigraph:rodriguez2007}. For example, scholars may need to:
\begin{itemize}\setlength{\itemsep}{-3pt}
	\item identify articles related to some interesting resource,
	\item	identify collaborators for a funding opportunity,
	\item identify a publication venue for a newly created article,
	\item identify referees to review an article, and
	\item identify resources of interest in one's community.
\end{itemize}

This article presents a context-sensitive recommender system to support the scholarly communication process. This system is called kReef.\footnote{kReef is currently available at \ttt{http://k-reef.com}.} kReef maintains a resource-rich, graph-based model of the scholarly community that includes people, articles, journals, conferences, calls, funding opportunities, institutions, etc. and their various relationships to one another. kReef utilizes this model to execute algorithms that codify problem-solving strategies in order to support scholars in various contexts. This article provides an overview of the kReef system from its data structures and algorithms to its user interface. 

\section{System Architecture}\label{sec:sysarch}

\begin{figure}[h!]
	\centering
	\includegraphics[width=0.325\textwidth]{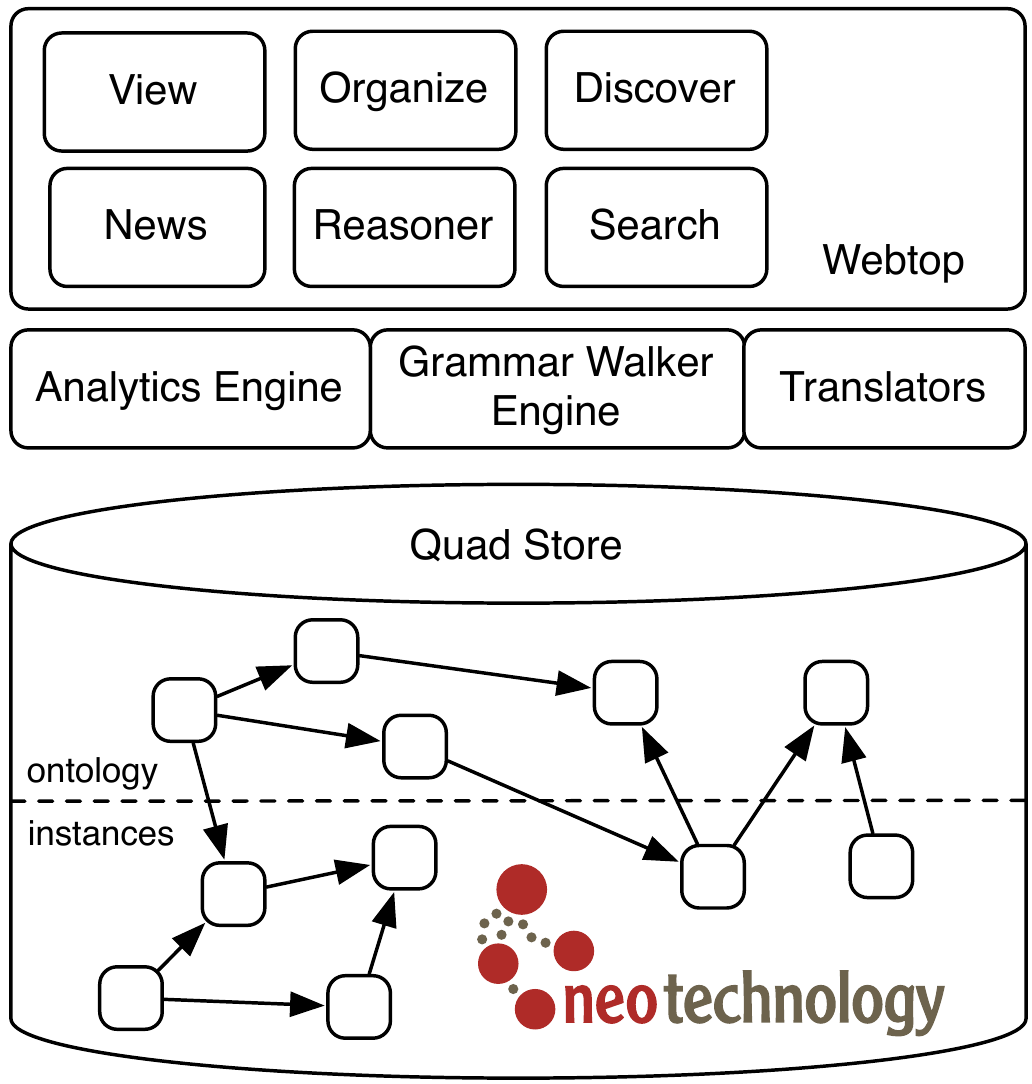}
	 \caption{\label{fig:stack} The subset of the kReef system to be discussed in this article ranging from data storage to user applications.}
\end{figure}

The foundation of kReef is a rich model of the scholarly community that includes various resource types and their relationships to one another. It was determined that the most appropriate standards and technologies to support such modeling are those set forth by the Semantic Web initiative \cite{lee:semantic2001}. Thus, kReef can be considered a Semantic Web technology. Figure \ref{fig:stack} diagrams those kReef components that will be discussed in this article.

\begin{figure}[h!]
	\centering
	\includegraphics[width=0.2\textwidth]{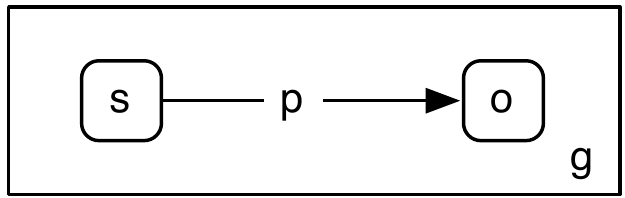}
	 \caption{\label{fig:quad} A quad is an $\la s, p, o, g \ra \in G$.}
\end{figure}

\section{The Quad Store}\label{sec:quadstore}

kReef's abstract data model is the quad-based representation of the Resource Description Framework (RDF) \cite{rdfcon:klyne2004,named:carroll2005}. If $U$ is the set of all Uniform Resource Identifiers (URI), $B$ is the set of all blank/anonymous nodes, and $L$ is the set of all literals, then a quad-based RDF graph is defined as
\begin{equation*}
	G \subseteq (U \cup B) \times U \times (U \cup B \cup L) \times (U \cup B),
\end{equation*}
where any $\la s, p, o, g \ra \in G$ is called a quad (or quad-based statement). The element $s$ is the subject, $p$ the predicate, $o$ the object, and $g$ the graph (or context). It is the role of $g$ to partition triple-based RDF statements of the form $\la s, p, o \ra$. Thus, two RDF statements that maintain the same $g$ are considered in the same graph. Figure \ref{fig:quad} diagrams an RDF quad.

An RDF quad store is a graph database that supports the representation and manipulation of RDF quads. Prior to the development and popularization of the quad concept, RDF stores were triple stores. Triple stores only support the $\la s, p, o\ra$ construct. However, it has become apparent that statement metadata (i.e.~reification) can be more efficiently represented using quads as opposed to the more verbose \ttt{rdf:Statement} construct \cite{named:carroll2005}. In kReef, the $g$ component of a quad serves as an identifier denoting a specific user's graph. Every time a user makes a statement, that statement is written to their graph---to their $g$. Currently, on the front-end, every user has a single graph unique to their account. In the future, users can have multiple graphs with different permissions. These permissions are controlled using Access Control List (ACL) metadata \cite{policy:reddivari2005}.

kReef uses Neo Technology's Neo4j graph database as its quad store.\footnote{Neo4j is currently available at \ttt{http://neo4j.org/}.} Neo4j makes use of a linked graph data structure in order to ensure optimal performance in graph traversals (i.e.~vertices have direct pointer's to their adjacent vertices). As will be explained later in the article, kReef reasoning and recommendation is not accomplished through typical monotonic description logics and reasoners as popularized and proselytized by the Semantic Web community \cite{baader:dl2003}. Instead, kReef performs reasoning and recommendation by means of the random walk algorithms popularized by the graph/network analysis community \cite{netanal:brandes2005}. For this reason, a quad store architecture that is optimized for traversals is a necessary requirement of kReef. Moreover, given the size of the scholarly community, a quad store that can scale to multiple billions of statements is another requirement. Neo4j meets both these needs.

\section{A Scholarly Ontology}\label{sec:ontology}

As previously discussed, the abstract data model for kReef is a quad-based RDF graph. The ontologies (or schemas) to structure the data are discussed in this section. kReef currently maintains two ontologies: \ttt{core}\footnote{The \ttt{core} namespace prefix resolves to \ttt{http://knowledgereefsystems.com/2007/11/core\#}.} and \ttt{relation}\footnote{The \ttt{relation} namespace prefix resolves to \ttt{http://knowledgereefsystems.com/2008/02/relation\#}.}. The \ttt{core} ontology represents objectively determinable resource and relationship types found in the scholarly community (see \S \ref{sec:core}). The \ttt{relation} ontology provides a set of predicates that are more subjective in nature (see \S \ref{sec:relation}).

\subsection{The Core Ontology}\label{sec:core}

\begin{figure*}
	\centering
	\includegraphics[width=0.85\textwidth]{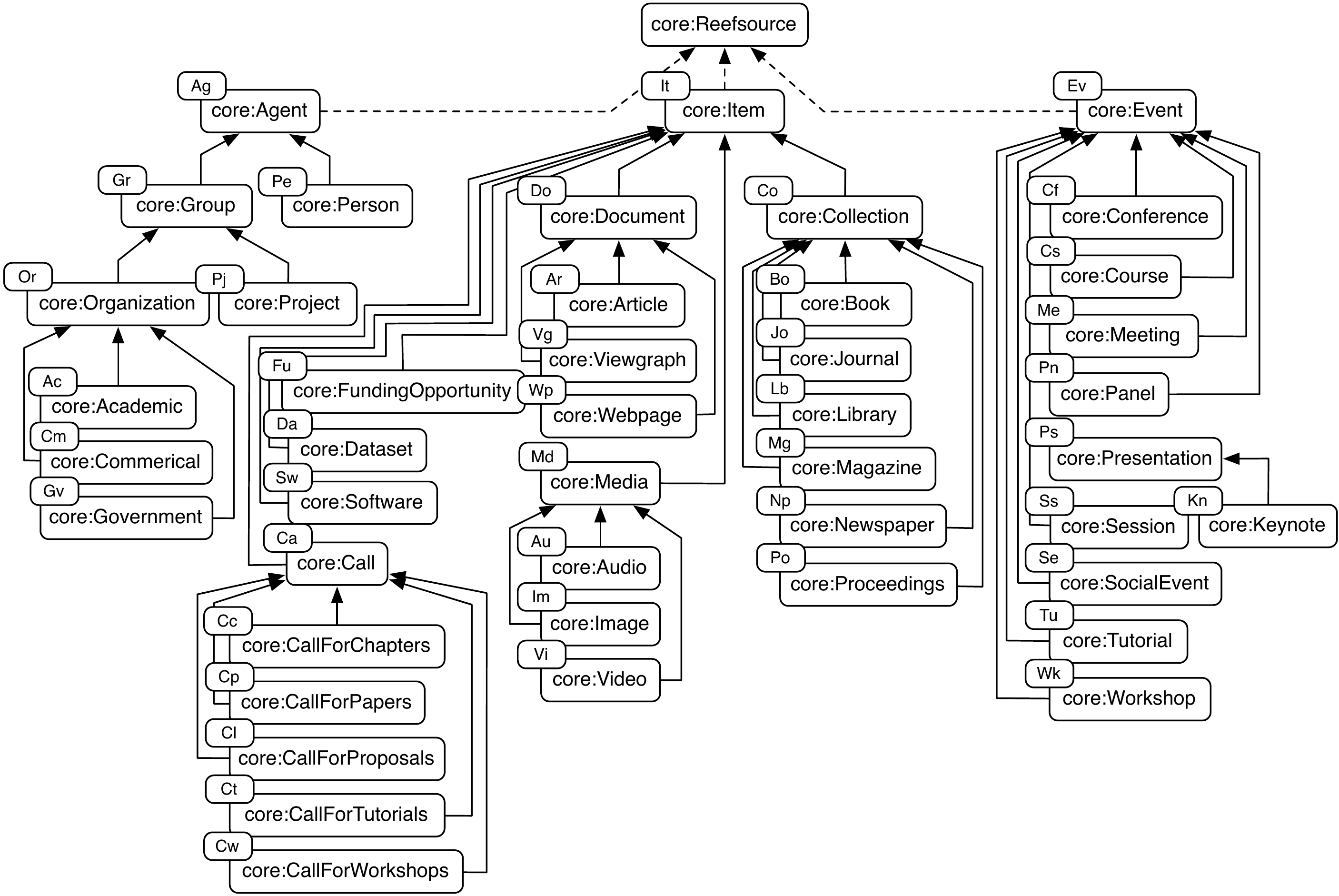}
	 \caption{\label{fig:core} The \ttt{rdfs:subClassOf} hierarchy of the \ttt{core} ontology. The solid lines represent explicit \ttt{rdfs:subClassOf} relations and the dashed lines represent implicit \ttt{rdfs:subClassOf} relations. Each \ttt{owl:Class} has a two character ``chemical" abbreviation used to visually denote a resource's \ttt{rdf:type} in the webtop user interface (see \S \ref{sec:webtop}).}
\end{figure*}

The \ttt{core} ontology is represented in the Web Ontology Language (OWL) \cite{owlspec:mcguinness2004}. OWL is a web-based, description logic language used to define class descriptions in order to infer which resource instances are subsumed by which classes. In kReef, the \ttt{core} ontology primarily serves as a schema to aid in data modeling.\footnote{Currently in kReef, ontologies are primarily used as schemas and not for description logic reasoning. The only reasoning rules currently supported by kReef are the \ttt{rdfs:subClassOf} and \ttt{rdfs:subPropertyOf} rules of the RDF Schema (RDFS) ruleset. The primary reason for this is that the computational complexity of description logic reasoning is high and it is necessary to ensure that kReef is a real-time system. However, into the future, kReef will support user generated \ttt{owl:Class} descriptions.} The form of the \ttt{core} ontology was inspired by the MESUR ontology \cite{onto:rodriguez2007} of the MESUR project \cite{bollen:mesur2008}. Figure \ref{fig:core} diagrams the \ttt{rdfs:subClassOf} hierarchy of \ttt{core}, where \ttt{core:Reefsource} is a direct \ttt{rdfs:subClassOf} of \ttt{owl:Thing}. For the \ttt{core:Event} branch, Jennifer Golbeck's \ttt{WWW04photo} ontology provided inspiration.\footnote{WWW04photo ontology is currently available at \ttt{http://www.mindswap.org/$\sim$golbeck/web/www04photo.owl}.} Note that not all \ttt{owl:Class}es are presented in Figure \ref{fig:core}. Some of the \ttt{owl:Class}es that are not represented pertain to support \ttt{owl:Class}es for the backend (e.g.~user account data, quad store data management, access control lists, etc.).

The \ttt{core} ontology maintains many \ttt{owl:ObjectProperty} and \ttt{owl:DatatypeProperty} relations. The most frequent ones are presented for each of the primary components of the \ttt{core} ontology: \ttt{core:Reefsource} (see Table \ref{tab:reefsource}), \ttt{core:Agent} (see Table \ref{tab:agent}), \ttt{core:Item} (see Table \ref{tab:item}), and \ttt{core:Event} (see Table \ref{tab:event}). For the sake of brevity, \ttt{owl:InverseProperty} and \ttt{owl:Restriction} information is not presented.

\begin{table}[h!]
\centering
\caption{\ttt{core:Reefsource} \ttt{rdf:Property} relations\label{tab:reefsource}}
\begin{scriptsize}
\begin{tabular}{|c|c|c|c|} \hline
\ttt{rdf:Property}&\ttt{rdfs:domain}&\ttt{rdfs:range}\\ \hline\hline
\ttt{core:title}&\ttt{core:Reefsource}&\ttt{xsd:string} \\ \hline
\ttt{core:abstract}&\ttt{core:Reefsource}&\ttt{xsd:string} \\ \hline
\ttt{core:guid}&\ttt{core:Reefsource}&\ttt{xsd:string} \\ \hline
\end{tabular}
\end{scriptsize}
\end{table}

\begin{table}[h!]
\centering
\caption{\ttt{core:Agent} \ttt{rdf:Property} relations\label{tab:agent}}
\begin{scriptsize}
\begin{tabular}{|c|c|c|c|} \hline
\ttt{rdf:Property}&\ttt{rdfs:domain}&\ttt{rdfs:range}\\ \hline\hline
\ttt{core:attends}&\ttt{core:Agent}&\ttt{core:Event} \\ \hline
\ttt{core:created}&\ttt{core:Agent}&\ttt{core:Item} \\ \hline
\ttt{core:member}&\ttt{core:Group}&\ttt{core:Person} \\ \hline
\ttt{core:subGroup}&\ttt{core:Group}&\ttt{core:Group} \\ \hline
\ttt{core:firstName}&\ttt{core:Person}&\ttt{xsd:string} \\ \hline
\ttt{core:lastName}&\ttt{core:Person}&\ttt{xsd:string} \\ \hline
\ttt{core:occupation}&\ttt{core:Person}&\ttt{xsd:string} \\ \hline
\ttt{core:sex}&\ttt{core:Person}&\ttt{core:Gender} \\ \hline
\end{tabular}
\end{scriptsize}
\end{table}

\begin{table}[h!]
\centering
\caption{\ttt{core:Item} \ttt{rdf:Property} relations\label{tab:item}}
\begin{scriptsize}
\begin{tabular}{|c|c|c|c|} \hline
\ttt{rdf:Property}&\ttt{rdfs:domain}&\ttt{rdfs:range}\\ \hline\hline
\ttt{core:cites}&\ttt{core:Item}&\ttt{core:Item} \\ \hline
\ttt{core:containedIn}&\ttt{core:Item}&\ttt{core:Collection} \\ \hline
\ttt{core:creationTime}&\ttt{core:Item}&\ttt{xsd:dateTime} \\ \hline
\ttt{core:doi}&\ttt{core:Item}&\ttt{xsd:anyURI} \\ \hline
\ttt{core:publisher}&\ttt{core:Item}&\ttt{core:Group} \\ \hline
\ttt{core:dueDate}&\ttt{core:Call}&\ttt{xsd:dateTime} \\ \hline
\ttt{core:callFor}&\ttt{core:Call}&\ttt{core:Reefsource} \\ \hline
\ttt{core:contains}&\ttt{core:Collection}&\ttt{core:Item} \\ \hline
\ttt{core:editor}&\ttt{core:Collection}&\ttt{core:Agent} \\ \hline
\ttt{core:isbn}&\ttt{core:Collection}&\ttt{xsd:anyURI} \\ \hline
\ttt{core:issn}&\ttt{core:Collection}&\ttt{xsd:anyURI} \\ \hline
\ttt{core:oaipmh}&\ttt{core:Library}&\ttt{xsd:anyURI} \\ \hline
\ttt{core:startPage}&\ttt{core:Article}&\ttt{xsd:int} \\ \hline
\ttt{core:endPage}&\ttt{core:Article}&\ttt{xsd:int} \\ \hline
\ttt{core:number}&\ttt{core:Article}&\ttt{xsd:int} \\ \hline
\ttt{core:volume}&\ttt{core:Article}&\ttt{xsd:int} \\ \hline
\end{tabular}
\end{scriptsize}
\end{table}

\begin{table}[h!]
\centering
\caption{\ttt{core:Event} \ttt{rdf:Property} relations\label{tab:event}}
\begin{scriptsize}
\begin{tabular}{|c|c|c|c|} \hline
\ttt{rdf:Property}&\ttt{rdfs:domain}&\ttt{rdfs:range}\\ \hline\hline
\ttt{core:startTime}&\ttt{core:Event}&\ttt{xsd:dateTime} \\ \hline
\ttt{core:endTime}&\ttt{core:Event}&\ttt{xsd:dateTime} \\ \hline
\ttt{core:presents}&\ttt{core:Event}&\ttt{core:Item} \\ \hline
\ttt{core:organizedBy}&\ttt{core:Event}&\ttt{core:Agent} \\ \hline
\ttt{core:subEvent}&\ttt{core:Event}&\ttt{core:Event} \\ \hline
\end{tabular}
\end{scriptsize}
\end{table}
		
\subsection{The Relation Ontology}\label{sec:relation}

\begin{figure}[h!]
	\centering
	\includegraphics[width=0.485\textwidth]{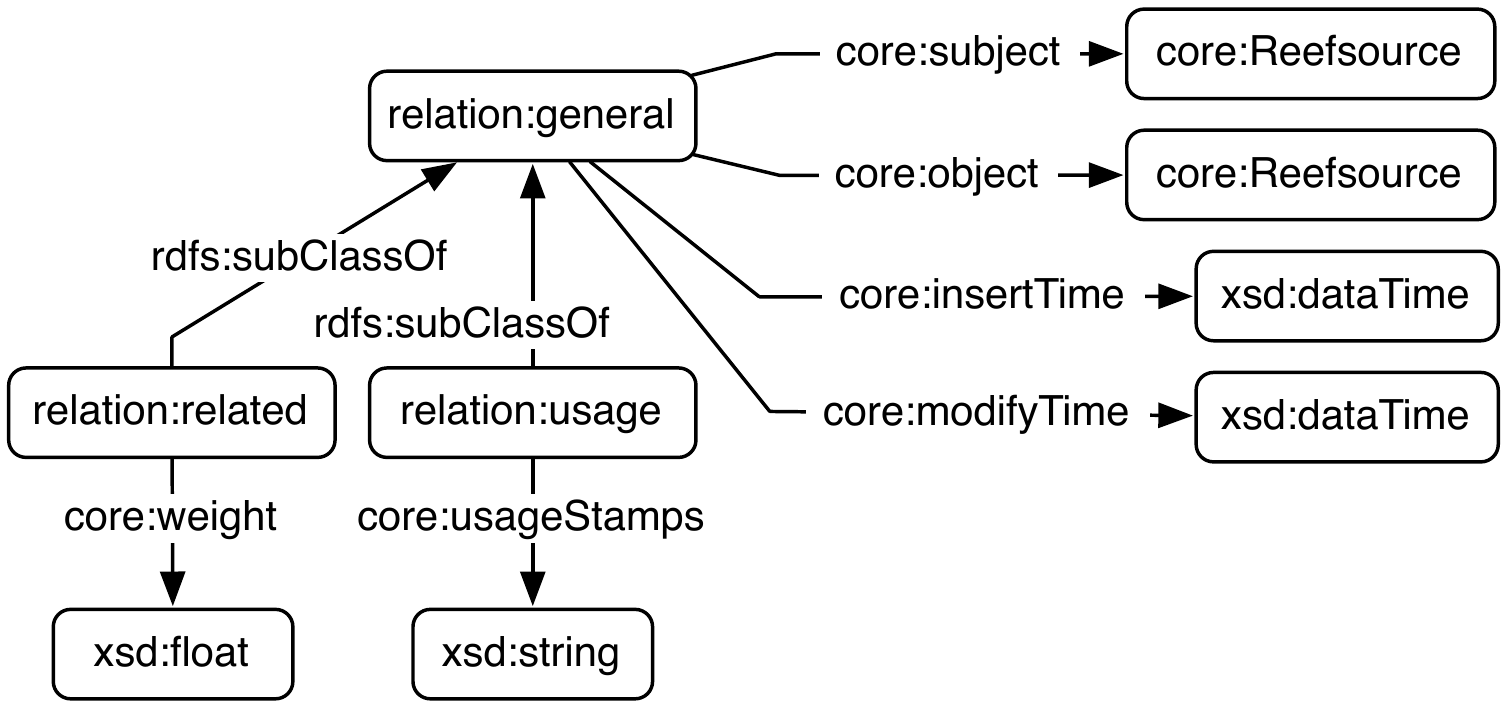}
	 \caption{\label{fig:relation}The \ttt{relation} ontology. Each \ttt{rdf:Property} is represented as a single edge, where the \ttt{rdfs:domain} is the tail of the edge and the \ttt{rdfs:range} is the head of the edge.}
\end{figure}

The purpose of the \ttt{relation} \ttt{owl:Class}es is to represent typed relationships between two resources and metadata about those relationships. In many ways, these \ttt{owl:Class}es are for reificiation. Unfortunately, given that $g$ is used to identify a user graph, to reify a statement with extra metadata, an \ttt{rdf:Statement}-like construct is required. Figure \ref{fig:relation} diagrams the \ttt{relation} \ttt{owl:Class}es, where the \ttt{rdfs:domain} and \ttt{rdfs:range} of their supported \ttt{rdf:Property} data are single directed edges.

The purpose of the \ttt{relation:related} \ttt{owl:Class} is to allow a user to explicitly denote their subjective interpretation of the similarity between two resources. Moreover, users can provide an \ttt{xsd:float} \ttt{core:weight} to denote the strength of similarity. This \ttt{core:weight} is a real number in $[0,1]$, where $0$ denotes a weak similarity and $1$ denotes a strong similarity. The most prevalent use of \ttt{related:relation} is to relate a \ttt{core:Concept} to a \ttt{core:Reefsource}. This is how users ``tag" resources. Given that a \ttt{core:Concept} is an \ttt{rdfs:subClassOf} \ttt{core:Reefsource}, users are also able to relate a \ttt{core:Concept} to another \ttt{core:Concept}. Thus, in the colloquial sense, users can ``tag tags". All created \ttt{relation:related} statements are written to the user's $g$ graph. This is how \ttt{relation:related} resources are tied to their creators. The role of \ttt{relation:related} is presented in \S \ref{sec:organize} and \S \ref{sec:news}.

The purpose of \ttt{relation:usage} is to track the ``resource path" that a user takes through kReef. When a user views one resource and then another, a \ttt{relation:usage} resource is created. For example, if at time step $t=1$, a user views resource $i$ and then at $t=2$ the user views resource $j$, then a \ttt{relation:usage} resource is created (if one does not already exist between $i$ and $j$ in that user's $g$). This \ttt{relation:usage} resource denotes $i$ as its \ttt{core:subject} and $j$ as its \ttt{core:object}. Moreover, time stamp $t=2$, which is represented as an \ttt{xsd:dateTime}, is turned into an \ttt{xsd:string} and recorded as the \ttt{rdfs:range} of the \ttt{core:usageStamps}. If the \ttt{relation:usage} resource between $i$ and $j$ already exists for that user, then the $t=2$ time stamp value is appended to the already existing \ttt{core:usageStamps} list. Again, because each $g$ is associated with a user, it is possible to determine at which time a particular user moved from resource $i$ to resource $j$.

\section{Translators}

The ontologies defined in \S \ref{sec:ontology} provide an abstract model of the scholarly community. In order to be useful, this abstract model must be populated with instances. The translators component serves this purpose. There are numerous translators that continually harvest scholarly data and represent it according to the previously presented ontologies. Fortunately, the scholarly community maintains a massive digital footprint that is represented in various repositories worldwide. The Open Access Initiative's Protocol for Metadata Harvesting (OAI-PMH) of the digital library community is an excellent mechanism for harvesting scholarly metadata \cite{oaipmh:lagoze2004}. An example translation of an OAI-PMH feed is provided below for the arXiv record \ttt{oai:arXiv.org:0807.2466}.\footnote{This record has been doctored to improve readability and for the sake of brevity. Also, for more information on the arXiv pre-print repository, please visit \ttt{http://arxiv.org}.}

\begin{small}
\begin{verbatim}
<record>
  <header>
    <identifier>oai:arXiv.org:0807.2466</identifier>
    <datestamp>2009-01-07</datestamp>
    <setSpec>cs</setSpec>
  </header>
  <metadata>
    <oai_dc:dc>
      <dc:title>A Grateful Dead Analysis...</dc:title>
      <dc:creator>Rodriguez, Marko A.</dc:creator>
      <dc:creator>Gintautas, Vadas</dc:creator>
      <dc:creator>Pepe, Alberto</dc:creator>
      <dc:subject>Computers and Society</dc:subject>
      <dc:subject>General Literature</dc:subject>
      <dc:subject>K.4.0</dc:subject>
      <dc:description>
        The Grateful Dead were an American band ...
      </dc:description>
      <dc:date>2008-07-15</dc:date>
      <dc:type>text</dc:type>
      <dc:identifier>
        http:// arxiv.org/abs/0807.2466
      </dc:identifier>
    </oai_dc:dc>
  </metadata>
</record>
\end{verbatim}
\end{small}

The record presented is represented in the Dublin Core (\ttt{dc}) schema.\footnote{The \ttt{dc} namespace prefix resolves to \ttt{http://purl.org/dc/elements/1.1/}.} There is a simple mapping from the \ttt{dc} schema to the \ttt{core} and \ttt{relation} ontologies.  Given this record, the arXiv translator will create a new \ttt{core:Article}. The \ttt{core:title} of the \ttt{core:Article} is determined by \ttt{dc:title}. The \ttt{core:abstract} of the \ttt{core:Article} is determined by \ttt{dc:description}. The \ttt{core:url} of the \ttt{core:Article} is determined by \ttt{dc:identifier}. The \ttt{core:guid} of the created \ttt{core:Article} is the identifier \ttt{oai:arXiv.org:0807.2466}. Next, three \ttt{core:Person} resources are created for the three \ttt{dc:creator}s and they are connected to the \ttt{core:Article} using \ttt{core:created}. The arXiv repository is considered a ``user" in kReef and thus, arXiv has its own $g$. In this way, it is possible to track which digital library repository provided which data. Moreover, given that arXiv has its own $g$, the \ttt{dc:subject} categories in each arXiv record is represented as a \ttt{relation:related} association between the \ttt{core:Article} and the \ttt{core:Concept} identified by \ttt{dc:subject}. This is how arXiv ``tags" its resources. Finally, various rules are used to ensure that duplicate resources are not created.

Other OAI-PMH feeds provide richer metadata such as journal and citation information. Also, other sources for scholarly metadata come from various RSS and ATOM feeds. More information about the kReef data providers can be found at \ttt{http://k-reef.com}. Finally, users are able to create and relate resources using the kReef webtop interface discussed in \S \ref{sec:webtop}. However, except for \ttt{relation:usage} data, most of the data in kReef comes from external repositories.

\section{Grammar Walker Engine}

In the current release of kReef, there are three recommender applications: Discover (see \S \ref{sec:discover}), Reasoner (see \S \ref{sec:reasoner}), and News (see \S \ref{sec:news}). The grammar walker engine is the kReef component that supports these applications. The engine executes functions similar in nature to constrained spreading activation \cite{inform:cohen1987} and the class of relative rank algorithms \cite{markov:white2003}. Gammar-based random walkers are random walkers designed specifically for multi-relational graphs \cite{grammar:rodriguez2008}. Most random walk techniques require a single-relational graph. However, an RDF graph is multi-relational as it supports multiple types of relationships between vertices (i.e.~there can be multiple predicates in an RDF graph). Along with some other process information, a grammar is an abstract path that a walker takes when traversing a multi-relational graph. The purpose of the algorithm is to identify resources related to some initial set of seed resources. The concept of ``relatedness" is determined by the grammar that the walker uses.

The grammar walker engine works as follows. Grammar walkers are initially distributed to a collection of seed resources and given a defined grammar. Next, they traverse the graph and increment counters for certain resource types they meet along the way as defined by their grammar.\footnote{In certain cases, especially for the Reasoner grammars, walkers can decrement counters.} Each walker has a time decaying ``energy" value $\epsilon \in \mbb{R}$ that they increment the resource counter with. The decay function governing the loss of walker energy is $\epsilon_{t+1} = \delta\epsilon_{t}$, where $\delta \in [0,1]$ is the decay parameter. When the walker energy decays below a given threshold or a certain number of prescribed steps have been taken, the process is complete. What is returned are the counters on the resources. This yields a ranked list that denotes those resources that are most related to the seed resources (according to the topology of $G$ and the grammar being used). Currently, all the recommender algorithms utilized in kReef make use of the grammar walker engine, where the grammar is tailored to particular problem-solving scenarios. In general, the success of graph-based methods for recommendation over correlation-based methods have been demonstrated in \cite{griffen:spread2006}.

\begin{figure}[h!]
	\centering
	\includegraphics[width=0.475\textwidth]{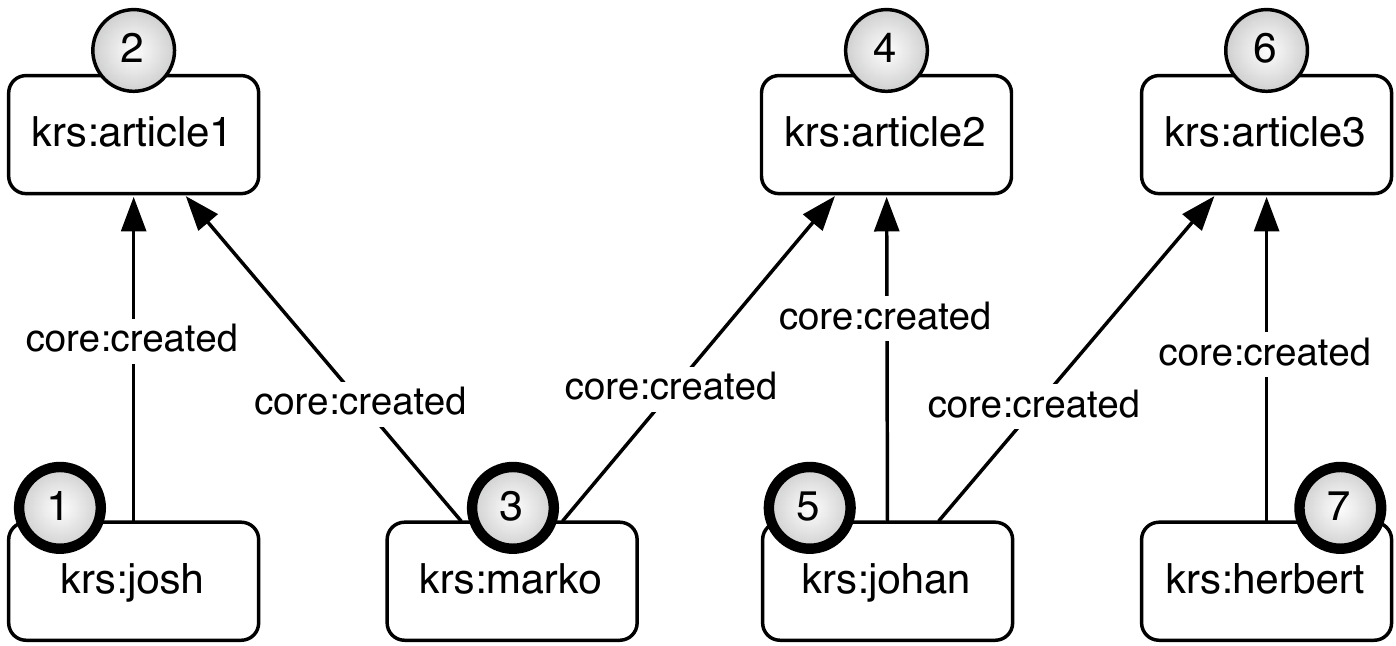}
	 \caption{\label{fig:coauthor-example} An example of a coauthorship grammar walk. There exists a single grammar walker denoted by the gray filled circle. The number in the walker denotes the time step at which the walker is at each resource. When the walker has a bold outline, the walker is updating the counter of its current resource. For the sake of diagram clarity, the  \ttt{core:createdBy} \ttt{owl:InverseProperty} is not presented.}
\end{figure}

A simple coauthorship grammar demonstrates the process. If a set of \ttt{core:Agent}s are needed that are ``coauthor-related" to some set of seed \ttt{core:Agent}s, then a coauthor grammar is used. Suppose a multi-relational graph with the following edge types and respective domains and ranges:
\begin{itemize}\setlength{\itemsep}{-3pt}
	\item $\ttt{core:created}: \ttt{core:Agent} \rar \ttt{core:Item}$
	\item $\ttt{core:createdBy}: \ttt{core:Item} \rar \ttt{core:Agent}$
\end{itemize}
Given these edge types only, there does not exist an explicit coauthorship graph. However, two \ttt{core:Agent}s are deemed coauthors if they have \ttt{core:created} the same \ttt{core:Item}. In order to traverse a coauthorship graph, a traversal of \ttt{core:created} and \ttt{core:createdBy} edges must be used. In order to ensure that a coauthorship graph is traversed, upon taking the \ttt{core:createdBy} edge, the grammar walker must not traverse to the same \ttt{core:Agent} it was located at on the previous step---as a \ttt{core:Agent} can not be their own coauthor. Also, only \ttt{core:Agent} resources have their counter incremented as the traversed \ttt{core:Item}s are not coauthor-related to the seed resources. Formally, if
\begin{eqnarray*}
\mca{A}^p_{s,o} =
	\begin{cases}
		1 & \text{if} \; \la s,p,o,* \ra \in G \\
		0 & \text{otherwise,}
	\end{cases}
\end{eqnarray*}
$\mbf{I}$ is the $\{0,1\}$-identity matrix, $p_1$ represents \ttt{core:created}, and $p_2$ represents \ttt{core:createdBy}, then the graph traversed by the grammar walker is defined as $(\mca{A}^{p_1} \cdot \mca{A}^{p_2}) \circ (\mbf{1} - \mbf{I})$.\footnote{The operator $\circ$ is the Hadamard entry-wise multiplication operation.} A diagram of this process is presented in Figure \ref{fig:coauthor-example}. Finally, an algebraic framework for representing any arbitrary grammar is presented in \cite{pathalg:rodriguez2008}.

\section{Analytics}\label{sec:analytics}

The analytics aspect of kReef provides statistics on resources. There are many ways to quantify the ``value" of a resource in the scholarly community: the Impact Factor \cite{impactreivew:garfield1999}, the H-Index \cite{index:hirsh2005}, the Y-Factor \cite{journalstatus:bollen2006}, etc. Many of these statistics can be easily derived in a quad store using SPARQL queries \cite{onto:rodriguez2007}. Moreover, given the amount of usage data that kReef records (i.e.~\ttt{relation:usage}), it is possible to provide statistics on how resources are used \cite{bollen:mesur2008}. The benefit of the analytics component is that it provides a user a quantified understanding of the impact of various resources in the scholarly community.

\section{Webtop}\label{sec:webtop}

The kReef webtop provides a desktop look and feel within a typical web browser. A user can have multiple windows open, can drag and drop resources between windows, and can shrink windows to conserve screen real-estate. The webtop applications currently supported are itemized below, where the $\dagger$-annotated applications are those that provide recommendations:
\begin{itemize}\setlength{\itemsep}{-3pt}
	\item View: used to view a resource
	\item Discover$^\dagger$: used for generic  recommendations
	\item Reasoner$^\dagger$: used for context-specific recommendations
	\item Organize: used to bookmark resources
	\item News$^\dagger$: used to identify community interests
	\item Search: used for typical keyword search\footnote{Given the simplicity of Search, there will be no dedicated subsection. In brief, Search is analogous to standard keyword search and returns a ranked list of those resources that have a user specified keyword in their \ttt{core:title} or \ttt{core:abstract}.}
\end{itemize}

\subsection{View\label{sec:view}}

\begin{figure}[h!]
	\centering
	\includegraphics[width=0.475\textwidth]{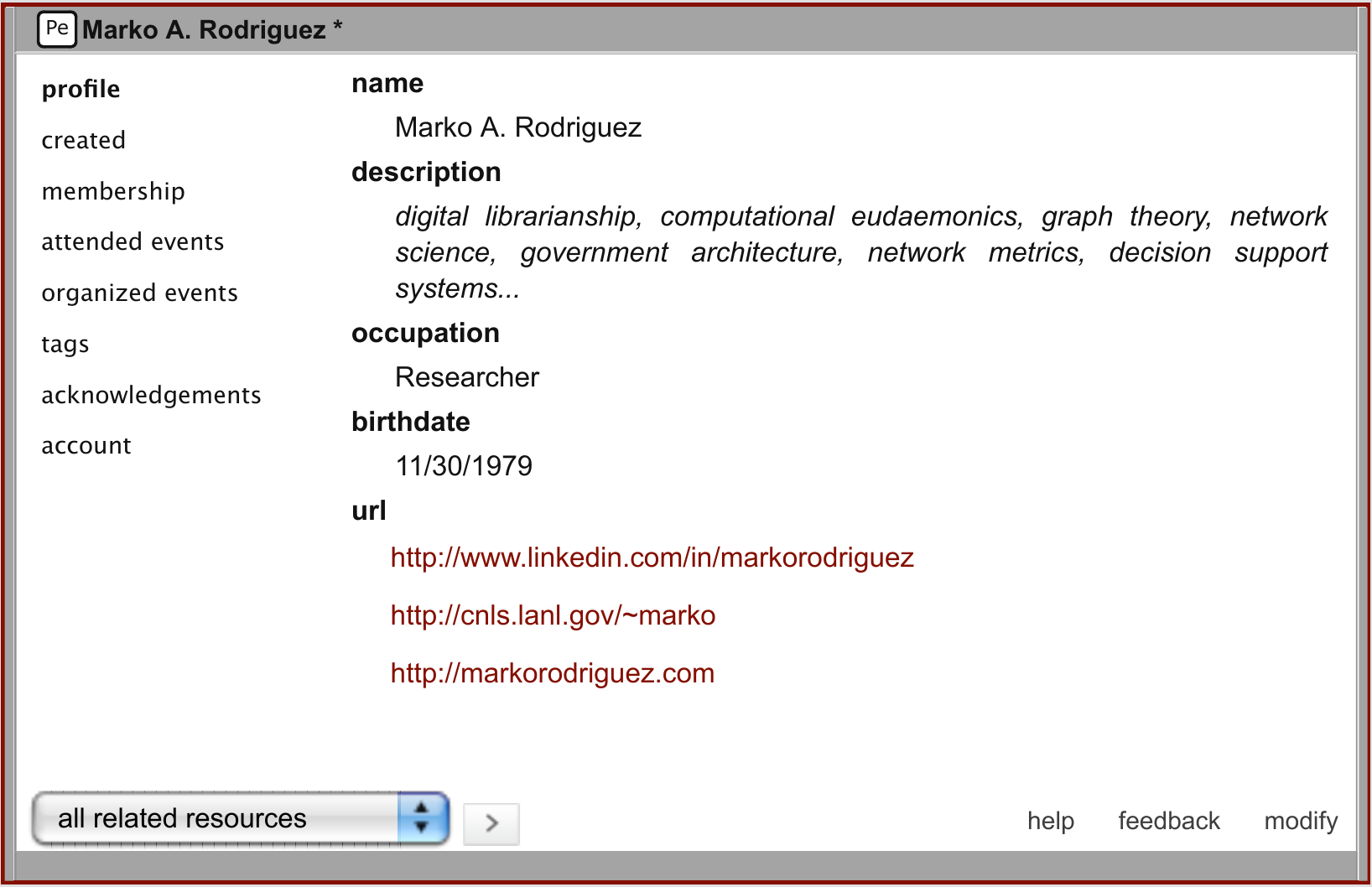}
	 \caption{\label{fig:view}A screenshot of a  \ttt{core:Person} View. The left hand side has various tabs. The right hand side provides the information contained in the currently selected tab.}
\end{figure}

View is analogous to a web browser, but instead of viewing HTML documents, the View renders subgraphs of the underlying RDF graph in a user-friendly manner. Each \ttt{core} \ttt{owl:Class} has its own specific rendering method. For instance, what data is gathered and rendered for a \ttt{core:Person} is different than the data that is gathered and rendered for a \ttt{core:Article}. Also, through View, users are able to modify and add information to a resource (i.e.~they can add \ttt{owl:DatatypeProperty} and \ttt{owl:ObjectProperty} information). Figure \ref{fig:view} provides a screenshot of a \ttt{core:Person} View that is rendering the \ttt{core:Person} resource with the \ttt{core:title} \ttt{"Marko A. Rodriguez"\^\!\^\!xsd:string}. Note that every \ttt{owl:Class} has an associated two character ``chemical" abbreviation. The abbreviation for \ttt{core:Person} is ``Pe". The abbreviations for each \ttt{owl:Class} are diagrammed in Figure \ref{fig:core}.

\subsection{Discover\label{sec:discover}}

\begin{figure}[h!]
	\centering
	\includegraphics[width=0.475\textwidth]{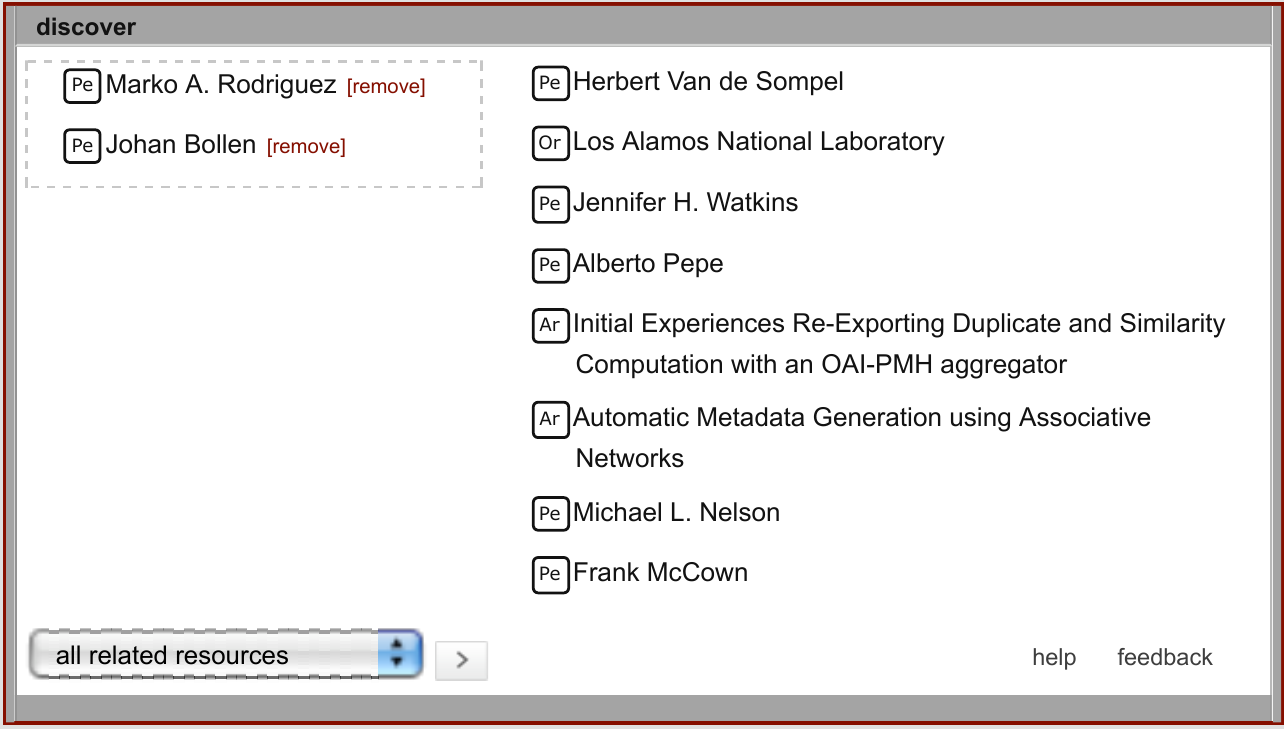}
	 \caption{\label{fig:discover}A screenshot of Discover. The left hand side has the user-provided seed resources. The right hand side has the recommended resources.}
\end{figure}

The Discover application is the simplest recommender application. Discover maintains a single general-purpose grammar that allows traversals along nearly all paths (except paths that will lead into the graph representation of the ontology). Furthermore, particular composite paths are accentuated such as coauthorship, co-citation, co-event, etc. In previous research, these composite paths have been deemed significant indicators of relatedness \cite{metadata:rodriguez2009}. Moreover, usage information is utilized in a collaborative filtering manner to identify co-used resources of interest \cite{video:bollen2007}. The only user settings in Discover are the seed resources and the desired return types (e.g.~\ttt{core:Agent}s, \ttt{core:Document}s, \ttt{core:Event}s, etc.). Figure \ref{fig:discover} presents a screenshot of Discover.

\subsection{Reasoner}\label{sec:reasoner}

More context-senstive recommendations are provided by the Reasoner application. Reasoner yields targeted solutions to particular scholarly problems. Current reasoning grammars can:
\begin{itemize}\setlength{\itemsep}{-1pt}
	\item provide initial resources with which to explore an idea,
	\item determine a \ttt{core:Group} of well suited collaborators,
	\item locate a \ttt{core:FundingOpportunity} to financially support an idea,
	\item locate an appropriate \ttt{core:Collection} or \ttt{core:Event} to which to submit a \ttt{core:Article}, and
	\item determine the most appropriate referees to review a \ttt{core:Article}.
\end{itemize}

Grammars for these particular problem-solving situations are currently designed according to intuition and have been validated using a small subset of test users.\footnote{Currently, approximately 20 users contribute to validating the Reasoner grammars.} The only reasoning grammar that has been rigorously validated through algorithmic techniques is the last grammar in the itemized list above: identify referees who are competent to review a submitted \ttt{core:Article}. Validation of this reasoner was originally presented in \cite{peeralg:rodriguez2006}, where the algorithm is able to make a statistically significant distinction between experts, non-experts, and conflict of interest referees. A description of this grammar is as follows. Given \ttt{core:Article} $i$, the user wants to identify a group of non-conflict of interest referees competent enough to review $i$. The authors of the \ttt{core:Article}s that are \ttt{core:citedBy} $i$ provide a set of competent referees. Moreover, the coauthors of the cited authors are also competent referees (recursively with decaying $\delta$ significance). However, authors tend to cite themselves and their collaborators. Thus, to remove conflict of interest referees, authors and the author's coauthors should be removed (recursively with decaying $\delta$). The ranked list returned provides a collection of expert referees for article $i$.

\subsection{Organize\label{sec:organize}}

Organize is analogous to a file system, but instead of organizing files, users organize resources. When a user creates a \ttt{relation:related} resource associating a \ttt{core:Concept} to a \ttt{core:Reefsource}, the \ttt{core:Concept} serves the function of a ``folder" and the \ttt{core:Reefsource} serves the function of a ``file" in that folder. Organize can also be thought of as a bookmark application where users can save resources they have found in kReef for later use. However, beyond bookmarking, Organize plays a significant role in the News recommendation application (see \S \ref{sec:news}). 


\subsection{News}\label{sec:news}

\begin{figure}[h!]
	\centering
	\includegraphics[width=0.475\textwidth]{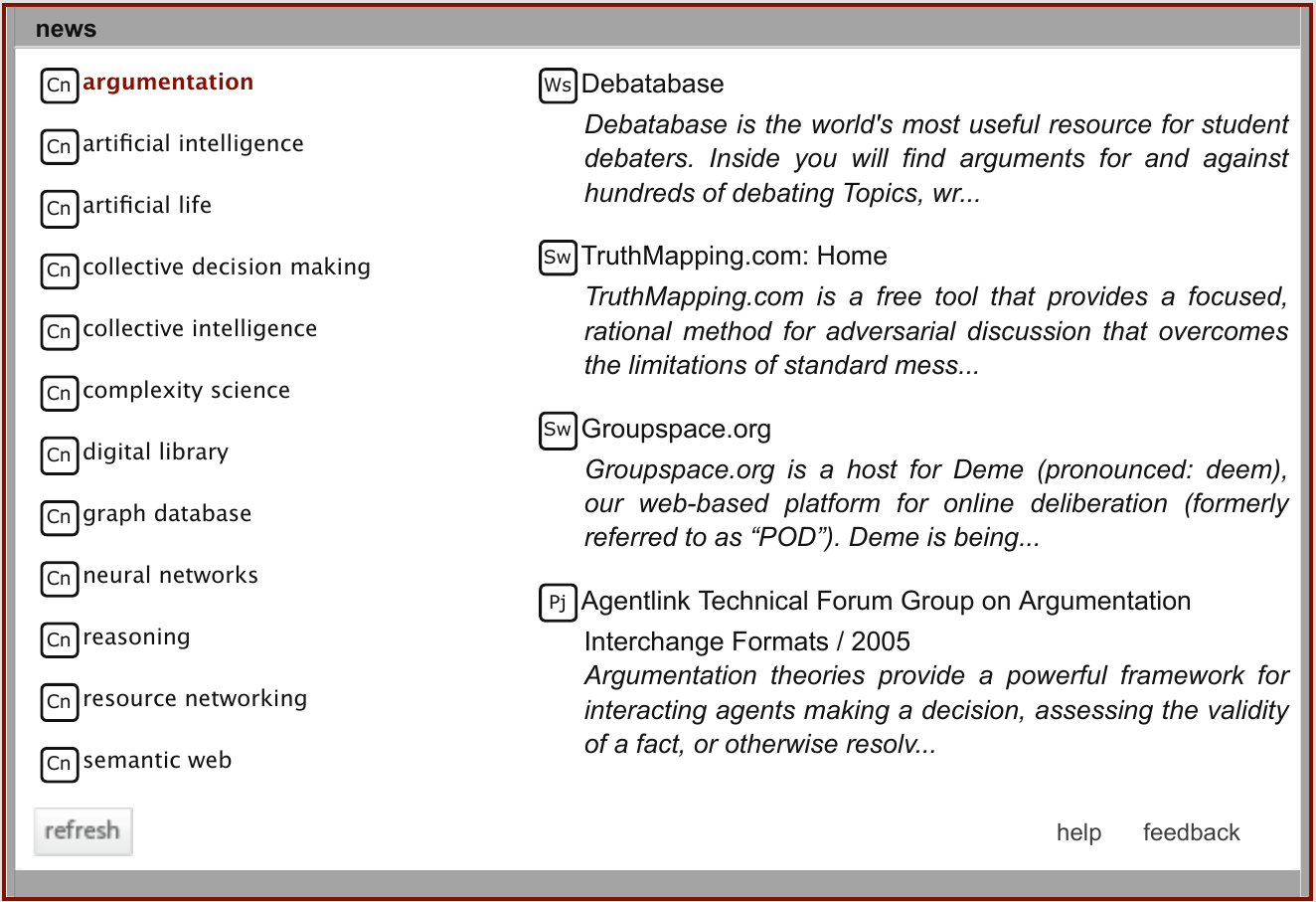}
	 \caption{\label{fig:news}A screenshot of the News application. The left hand side denotes the user's \ttt{core:Concept}s and the right hand side denotes the various resources that his respected/trusted \ttt{core:Agent}s have tagged according to those \ttt{core:Concept}s.}
\end{figure}

There is no such thing as an explicit ``social network" in kReef. For one, there is no notion of ``friendship". Users, like other scholarly artifacts, are resources. However, a user is a multi-faceted entity that can be considered a resource for various ends. For example, when user $i$ tags user $j$ with \ttt{core:Concept} $k$, user $i$ is stating that they respect/trust user $j$ in the area of \ttt{core:Concept} $k$. Thus, the social graph that emerges is a multi-relational graph denoting how users respect each other. Moreover, this multi-relational social graph serves as a medium for propagating targeted resources between users. The purpose of the News application is to allow users to view this propagation. In this way, News serves as a type of context-sensitive RSS feed which recommends resources of interest in the user's community. Figure \ref{fig:news} provides a screenshot of the News application.

All of the \ttt{core:Concept}s that a user has in their Organize application can be seen on the left hand side of the News application. When user $i$ clicks on \ttt{core:Concept} $k$, user $i$ diffuses a swarm of News grammar walkers. The grammar walkers are seeded on all the users that user $i$ has tagged (i.e.~\ttt{relation:related}) according to \ttt{core:Concept} $k$. The walkers then traverse to all the resources that this set of users have tagged as $k$. If one of those resources is yet another user, then the process repeats. Each time a walker traverses an edge, it decays its energy according to the exponential decay function $\epsilon_{t+1} = 2^{-\Delta / \sigma}\epsilon_{t}$, where $\Delta$ is the difference in time between the current time and the \ttt{core:insertTime} of the \ttt{relation:related} resource and $\sigma$ is the half-life of the significance of the association. The exponential decay function ensures that those resources that have been more recently \ttt{relation:related} are more highly recommended. Ultimately, what is returned is a ranked list of those resources tagged as $k$ that are topologically near user $i$ in the multi-relational social graph generated by $k$.

\begin{figure}[h!]
	\centering
	\includegraphics[width=0.475\textwidth]{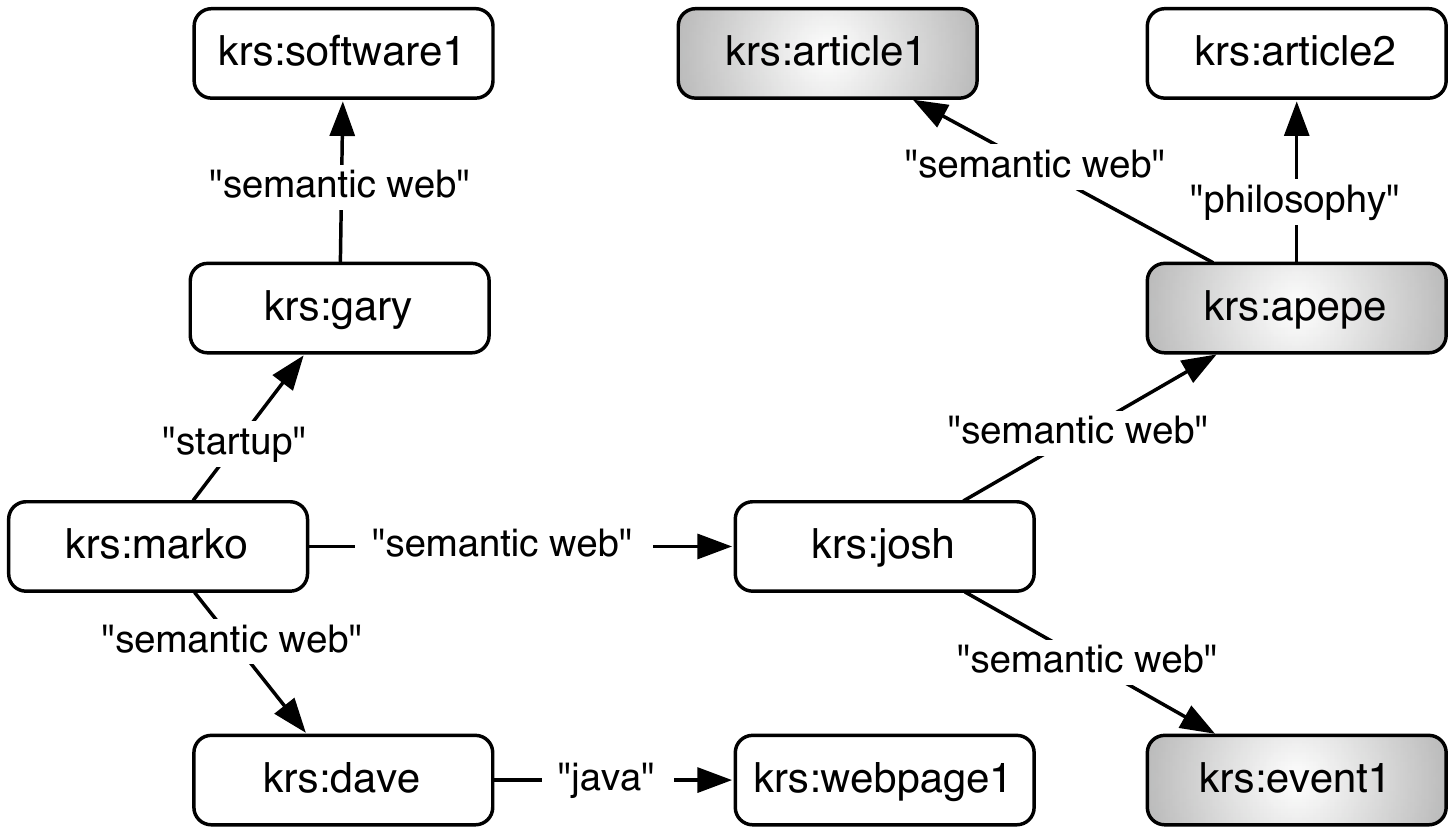}
	 \caption{\label{fig:news-example}An example ``semantic web" News recommendation for the \ttt{krs:marko} user. The highlighted resources are the recommended resources.}
\end{figure}

Figure \ref{fig:news-example} provides and example of how News works. This example is from the perspective of user \ttt{krs:marko} and according to the ``semantic web" \ttt{core:Concept}. If \ttt{krs:marko} wants to get recommended ``semantic web" resources in his News application, a grammar is executed that follows ``semantic web" \ttt{relation:related} edges from \ttt{krs:marko}.\footnote{Realize that these are not explicit statements in the quad store as \ttt{relation:related} is a collection of statements.} Given that \ttt{krs:marko} does not respect \ttt{krs:gary} according to ``semantic web", he does not see \ttt{krs:gary}'s \ttt{krs:software1} resource. Even though \ttt{krs:marko} respects \ttt{krs:dave} in terms of ``semantic web", he does not see \ttt{krs:dave}'s \ttt{krs:webpage1} recommendation as it is tagged ``java". Given that \ttt{krs:josh} respects \ttt{krs:apepe} in terms of ``semantic web" and \ttt{krs:apepe} respects \ttt{krs:article1} in terms of ``semantic web", \ttt{krs:marko} sees both \ttt{krs:apepe} and \ttt{krs:article1} in his ``semantic web" recommendation. Note that \ttt{krs:dave} and \ttt{krs:josh} are not recommended to \ttt{krs:marko} because these resources are already explicitly respected by \ttt{krs:marko}.

\section{Conclusion}

kReef is a service that supports the scholarly communication process. Many of the standards and technologies of the Semantic Web initiative have been incorporated into its design and implementation. Additionally, kReef takes advantage of the graph structure of its data set by applying techniques from the network analysis community. Among these techniques is a general-purpose framework for performing random walks in a multi-relational graph. The grammar walker engine currently serves as the core recommendation technology in kReef. This engine derives intuitive, personalized, context-sensitive recommendations from harvested and user-generated data. The webtop user interface delivers these recommendations to users in real time. The future of kReef will include not only more algorithms tailored to scholarly contexts, but also an infrastructure to support virtual and real-time scholarly communication that is believed to be more efficient and effective than the scholarly communication infrastructure present today.

\section{Acknowledgements}

Many people have worked on the kReef project to varying degrees. The following people are acknowledged in order of their level of contribution: Seth Horne, Kerri Korschgen, John Hellier, Carl Sylvia, and Brian DeSpain. Jennifer H. Watkins of the Los Alamos National Laboratory provided editorial assistance on the article. The Neo Technology team provided much assistance with the backend aspects of kReef. The News recommender algorithm was developed by Marko A. Rodriguez, Daniel Steinbock, and Jennifer H. Watkins of the Los Alamos National Laboratory and Stanford University. This algorithm is currently patent pending with the U.S. Patent and Trademark Office under application number 11/608,586. For more information on kReef, please visit \ttt{http://k-reef.com}.

\balancecolumns

\begin{thebibliography}{10}

\bibitem{baader:dl2003}
F.~Baader, D.~Calvanese, D.~L. Mcguinness, D.~Nardi, and P.~F. Patel-Schneider,
  editors.
\newblock {\em The Description Logic Handbook: Theory, Implementation and
  Applications}.
\newblock Cambridge University Press, January 2003.

\bibitem{lee:semantic2001}
T.~Berners-Lee, J.~A. Hendler, and O.~Lassila.
\newblock The {S}emantic {W}eb.
\newblock {\em Scientific American}, pages 34--43, May 2001.

\bibitem{video:bollen2007}
J.~Bollen, M.~L. Nelson, G.~Geisler, and R.~Araujo.
\newblock Usage derived recommendations for a video digital library.
\newblock {\em Journal of Network and Computer Applications}, 30(3):1059--1083,
  2007.

\bibitem{journalstatus:bollen2006}
J.~Bollen, M.~A. Rodriguez, and H.~{Van de Sompel}.
\newblock Journal status.
\newblock {\em Scientometrics}, 69(3):669--687, December 2006.

\bibitem{bollen:mesur2008}
J.~Bollen, H.~{Van de Sompel}, and M.~A. Rodriguez.
\newblock Towards usage-based impact metrics: first results from the {MESUR}
  project.
\newblock In {\em {Proceedings of the Joint Conference on Digital Libraries}},
  pages 231--240, New York, NY, 2008. {IEEE/ACM}.

\bibitem{netanal:brandes2005}
U.~Brandes and T.~Erlebach, editors.
\newblock {\em Network Analysis: Methodolgical Foundations}.
\newblock Springer, Berling, DE, 2005.

\bibitem{named:carroll2005}
J.~J. Carroll, C.~Bizer, P.~Hayes, and P.~Stickler.
\newblock Named graphs, provenance and trust.
\newblock In {\em {Proceedings of the International World Wide Web
  Conference}}, pages 613--622, Chiba, Japan, 2005. {ACM} Press.

\bibitem{inform:cohen1987}
P.~R. Cohen and R.~Kjeldsen.
\newblock Information retrieval by constrained spreading activation in semantic
  networks.
\newblock {\em Information Processing and Management}, 23(4):255--268, 1987.

\bibitem{impactreivew:garfield1999}
E.~Garfield.
\newblock Journal impact factor: a brief review.
\newblock {\em Canadian Medical Association Journal}, 161:979--980, 1999.

\bibitem{griffen:spread2006}
J.~Griffith, C.~O'Riordan, and H.~Sorensen.
\newblock {\em Knowledge-Based Intelligent Information and Engineering
  Systems}, volume 4253 of {\em Lecture Notes in Artificial Intelligence},
  chapter A Constrained Spreading Activation Approach to Collaborative
  Filtering, pages 766--773.
\newblock Springer-Verlag, 2006.

\bibitem{index:hirsh2005}
J.~Hirsch.
\newblock An index to quantify an individual's scientific output.
\newblock {\em Proceedings of the National Academy of Science},
  102(46):16569--16572, November 2005.

\bibitem{rdfcon:klyne2004}
G.~Klyne and J.~J. Carroll.
\newblock Resource description framework ({RDF}): Concepts and abstract syntax.
\newblock Technical report, World Wide Web Consortium, 2004.

\bibitem{oaipmh:lagoze2004}
C.~Lagoze, H.~{Van de Sompel}, M.~L. Nelson, and S.~Warner.
\newblock The {O}pen {A}rchives {I}nitiative {P}rotocol for {M}etadata
  {H}arvestings, October 2004.

\bibitem{owlspec:mcguinness2004}
D.~L. McGuinness and F.~van Harmelen.
\newblock {OWL} web ontology language overview, February 2004.

\bibitem{policy:reddivari2005}
P.~Reddivari, T.~Finin, and A.~Joshi.
\newblock Policy based access control for a {RDF} store.
\newblock In {\em {Proceedings of the International World Wide Web
  Conference}}, pages 78--83, Chiba, Japan, May 2005. {ACM} Press.

\bibitem{recommned:resnick1997}
P.~Resnick and H.~R. Varian.
\newblock Recommender systems.
\newblock {\em Communications of the {ACM}}, 40(3):56--58, 1997.

\bibitem{multigraph:rodriguez2007}
M.~A. Rodriguez.
\newblock A multi-relational network to support the scholarly communication
  process.
\newblock {\em International Journal of Public Information Systems},
  2007(1):13--29, 2007.

\bibitem{grammar:rodriguez2008}
M.~A. Rodriguez.
\newblock Grammar-based random walkers in semantic networks.
\newblock {\em Knowledge-Based Systems}, 21(7):727--739, 2008.

\bibitem{onto:rodriguez2007}
M.~A. Rodriguez, J.~Bollen, and H.~{Van de Sompel}.
\newblock A practical ontology for the large-scale modeling of scholarly
  artifacts and their usage.
\newblock In {\em {Proceedings of the Joint Conference on Digital Libraries}},
  pages 278--287, New York, NY, June 2007. {IEEE/ACM}.

\bibitem{peeralg:rodriguez2006}
M.~A. Rodriguez, J.~Bollen, and H.~{Van de Sompel}.
\newblock An algorithm to determine peer-reviewers.
\newblock In {\em {Proceedings of the Conference on Information and Knowledge
  Management}}, pages 319--328, Napa, California, October 2008. {ACM} Press.

\bibitem{metadata:rodriguez2009}
M.~A. Rodriguez, J.~Bollen, and H.~{Van de Sompel}.
\newblock Automatic metadata generation using associative networks.
\newblock {\em {ACM} Transactions on Information Systems}, 27(2):1--20, March
  2009.

\bibitem{pathalg:rodriguez2008}
M.~A. Rodriguez and J.~Shinavier.
\newblock Exposing multi-relational networks to single-relational network
  analysis algorithms.
\newblock Technical Report LA-UR-08-03931, Los Alamos National Laboratory,
  2008.

\bibitem{markov:white2003}
S.~White and P.~Smyth.
\newblock Algorithms for estimating relative importance in networks.
\newblock In {\em {Proceedings of the International Conference on Knowledge
  Discovery and Data Mining}}, pages 266--275, New York, NY, 2003. ACM Press.

\end{thebibliography}
\end{document}